# Valley Depolarization in Monolayer Transition-Metal Dichalcogenides with Zone-Corner Acoustic Phonons

Tae-Young Jeong,[†a,b] Soungmin Bae,[†c,d] Seong-Yeon Lee,[a] Suyong Jung,[b] Yong-Hoon Kim,[c,*] and Ki-Ju Yee[a,*]

## Abstract

**Although single-layer transition-metal dichalcogenides with novel valley functionalities are promising candidate to realize valleytronic devices, the essential understanding of valley depolarization mechanisms is still incomplete. Based on pump-probe experiments performed for MoSe$_2$ and WSe$_2$ monolayers and corroborating analysis from density functional calculations, we demonstrate that coherent phonons at the K-point of the Brillouin zone can effectively mediate the valley transfer of electron carriers. In the MoSe$_2$ monolayer case, we identify this mode as the flexural acoustic ZA(K) mode, which has broken inversion symmetry and thus can enable electron spin-flip during valley transfer. On the other hand, in the monolayer WSe$_2$ case where spin-preserving inter-valley relaxations are preferred coherent LA(K) phonons with even inversion symmetry are efficiently generated. These findings establish that, while the specifics of inter-valley relaxations depend on the spin alignments of energy bands, the K-point phonons should be taken into account as an effective valley depolarization pathway in transition metal dichalcogenide monolayers.**

## Introduction

Enormous exciton binding energies and the direct-indirect energy gap transition depending on layer number in two-dimensional transition metal dichalcogenides (TMDCs) have drawn extensive theoretical and experimental research interests over the last decade.[1-3] Specifically, type-VI semiconducting monolayer TMDC films, such as MoS$_2$, WS$_2$, MoSe$_2$, and WSe$_2$ with a trigonal prismatic (2H) crystalline structure, are spin polarized at different valleys because of strong spin-orbit coupling (SOC) and broken inversion symmetry.[2, 4, 5] The large SOC-induced splitting combined with the characteristic valley-spin locking is considered as a key material specification for semiconducting monolayer TMDC films to be useful in next-generation valleytronics applications.[6, 7]

Because valley functionalities heavily rely on the coherence times of the spins and valleys, the valley dynamics and scattering mechanisms are of critical importance in realizing valleytronic devices based on monolayer TMDC films. Spin-polarized photoluminescence[8-11] and time-resolved Kerr or Faraday rotation measurements [12-19] were generally adopted to investigate valley-dependent optical properties, and the exciton–exciton exchange interaction has been suggested as the primary mechanism leading to the temperature- and carrier density-dependent valley depolarization in monolayer MoSe$_2$.[20-22] However, alternatively, carrier-phonon interactions were claimed to be the main source that explains the excitation energy dependence of valley polarized photoluminescence.[23-27] Overall, despite the great research interest and effort in this topic, there remain important unanswered problems, especially regarding the underlying physical mechanisms of the valley dynamics and the inter-valley scattering.

Generation of coherent lattice vibrations or phonons could be a valuable tool for the study of phonon dynamics and carrier-phonon interactions in monolayer TMDCs, particularly near resonant excitation conditions where conventional Raman scatterings are hindered by the strong radiative recombination. Ultrashort optical pulses applied to an absorbing material can instantaneously activate coherent phonons (CPs), which are visualized by measuring the temporal modulation in the optical properties. Because the total momentum is conserved during the carrier–phonon scattering, the participating phonon should compensate for any possible momentum transfer of electrons. The temporal evolution of CPs and the underlying electron–phonon interactions can be investigated by performing time-resolved pump-probe experiments.[28-32]

[a.] Department of Physics, Chungnam National University, Daejeon 34134, Republic of Korea. E-mail: kyee@cnu.ac.kr
[b] Quantum Technology Institute, Korea Research Institute of Standards and Science, Daejeon 34113, Republic of Korea.
[c] School of Electrical Engineering, Korea Advanced Institute of Science and Technology (KAIST), Daejeon 34141, Korea. E-mail: y.h.kim@kaist.ac.kr
[d] Department of Physics, Yokohama National University, Hodogaya-ku, Yokohama 240-8501, Japan.

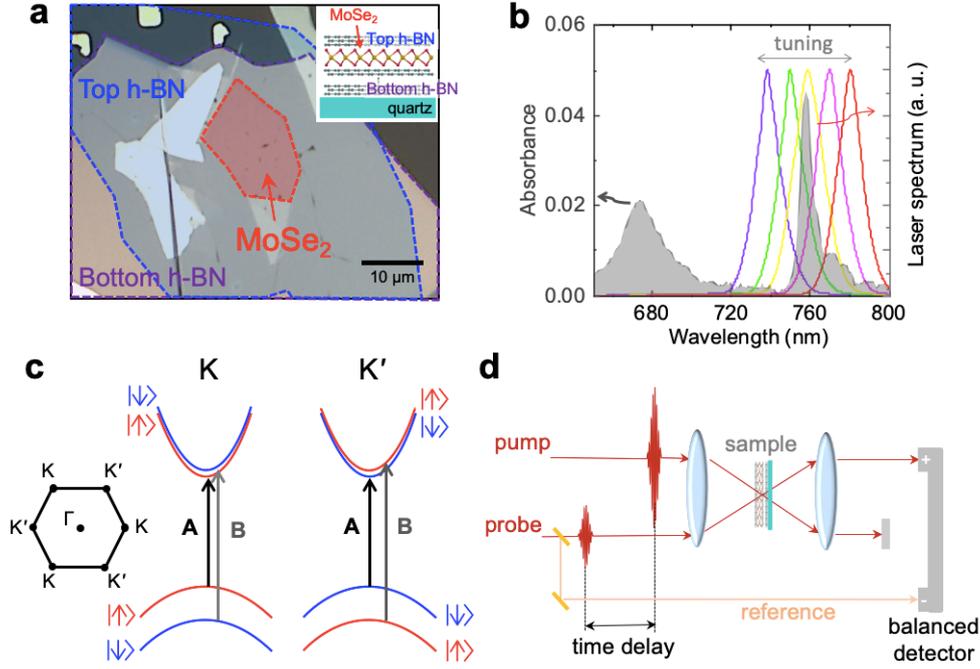

**Fig. 1** Monolayer MoSe$_2$ encapsulated by h-BN flakes and schematic pump-probe setup. (a) Optical microscope image of the monolayer MoSe$_2$ encapsulated by h-BN flakes at the top and the bottom. The purple and blue dash lines are, respectively, the boundaries of top and bottom h-BN, and the red line encloses the monolayer MoSe$_2$. The inset is a schematic image of the sample's side view. (b) Absorbance spectrum at a temperature of 80 K. Some of the laser spectra used in the pump-probe measurements are plotted as colored lines. (c) Energy bands diagram at the K- and K'-valleys of monolayer MoSe$_2$. (d) Schematic experimental setup of the pump-probe technique for measuring coherent phonons. Balanced detection is applied to improve the signal-to-noise ratio.

In this article, carrying out time-resolved pump-probe measurements and corroborating density functional theory (DFT) calculations on MoSe$_2$ and WSe$_2$ monolayers, we present hitherto undiscussed electron-phonon coupling-induced valley depolarization channels. Specifically, we show that the valley dynamics in these monolayer TMDCs are strongly affected by coherent acoustic phonons generated at the corners of the Brillouin zone. For the case of monolayer MoSe$_2$, we identify the CPs involved in the excited electron transfer from the K valley to the K' valley as the zone-corner flexural acoustic ZA(K) phonon mode. The requirement of electron spin-flip during this inter-valley scattering process explains according to the Eliott–Yafet mechanism the activation of only the out of plane flexural ZA(K) phonon-mode symmetry in monolayer MoSe$_2$. In contrast, in monolayer WSe$_2$ that has opposite valley-coupled spin state configurations, we identify that longitudinal acoustic LA(K) phonons are efficiently generated. The correlation between the mirror symmetry of phonon modes and the electron spin-flip process, as analyzed by our combined experiments and computations, reinforces the significant role of the K-point phonons in the valley depolarization of semiconducting monolayer TMDCs. The 2LA(K) or LA(K) phonon scattering was previously suggested as the valley depolarization mechanism in Mo-based TMDCs, [23-25] but to our knowledge the ZA(K) phonon-mediated channel is invoked for the first time. Moreover, the detailed comparison of the ZA(K) and LA(K) phonon activation processes in MoSe$_2$ and WSe$_2$, respectively, will shed new light on the valley depolarization mechanisms.

## Results and discussion

Fig. 1a shows an optical microscope image of monolayer MoSe$_2$ encapsulated with high-quality h-BN flakes on a quartz substrate. Purple and blue dashed lines respectively indicate the bottom and top h-BN, and the red dashed line outlines the monolayer MoSe$_2$. Encapsulation of monolayer MoSe$_2$ with h-BN layers not only reduces defect formations from unwanted molecular adsorptions but also minimizes disorders from surface roughness and charged dopants from the substrates, thus resulting in a very narrow exciton linewidth[33, 34]. The absorbance spectrum in Fig. 1b obtained from the encapsulated monolayer MoSe$_2$ at 80 K indicates A-exciton resonance at ≈ 758 nm with a full-width at half-maximum linewidth of ≈ 6 nm. As outlined in Fig. 1c, the energetic ordering of electron spins at K-valleys is opposite to that at K' valleys, and the energy splitting is particularly large in the valence bands. Accordingly. in addition to the A-exciton signal, we observe relatively broad B-exciton absorbance at ≈ 674 nm, from which we can deduce the valley-coupled spin splitting of about 200 meV.

In order to investigate CP generation and valley dynamics, we carried out time-resolved transmittance measurements using ultrashort optical pulses with a duration time of ≈ 80 fs generated with a mode-locked Ti:sapphire oscillator. As indicated in Fig. 1b, the excitation wavelength was tuned around the A-exciton resonance, from 740 to 780 nm. In Fig. 1d we show the schematic of the pump-probe experimental setup for CP measurements. Pump pulses instantaneously create CPs as well as non-equilibrium charged carriers; then, the temporal evolution of the CPs and the photo-excited carriers can be obtained from the transmittance change (ΔT) of the probe pulse as a function of time delay relative to the pump pulse. Here, a balanced detection scheme with a reference beam was adopted to reduce signal noise by compensating for laser fluctuation.

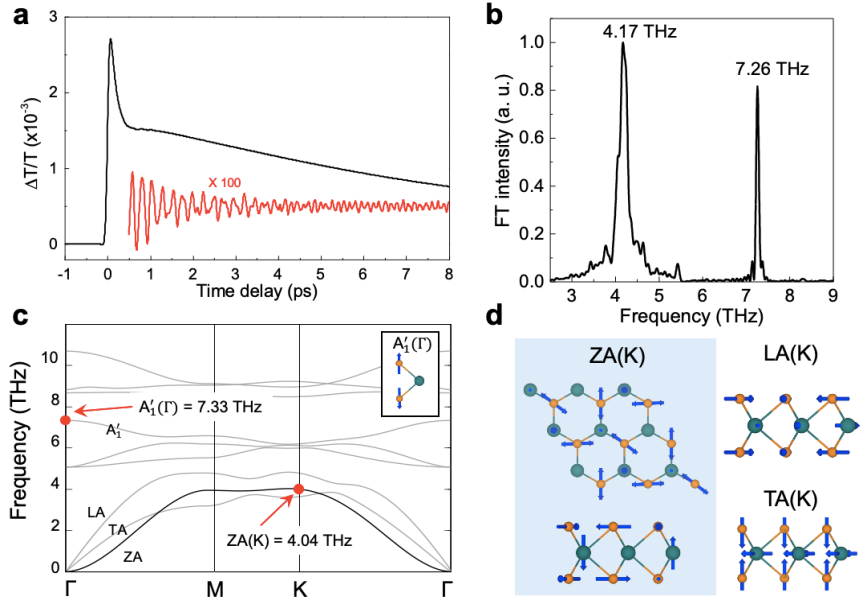

**Fig. 2** Coherent phonon oscillations in MoSe$_2$ monolayer. (a) Time-resolved transmission change in monolayer MoSe$_2$ obtained with an excitation energy of 755 nm. The red line plots the magnified transmission oscillation due to CP generation. (b) Fourier transformation spectrum of the ΔT oscillation, exhibiting two phonon modes at 4.17 THz and 7.26 THz. (c) Phonon dispersion of monolayer MoSe$_2$ calculated with the density functional theory. The red dots indicate the Γ-point A'$_1$ (Γ) mode at 7.33 THz and ZA(K) at 4.04 THz. The inset depicts the atomic motion of the A'$_1$(Γ) mode. d, Atomic motions of three K-point acoustic modes ZA(L), LA(K), and TA(K).

As is shown in Fig. 2a, the time-resolved transmission change (ΔT/T) obtained at a wavelength of 755 nm exhibits bi-exponential decay of the signal created at zero time delay by the pump pulse. We designate the fast decay signal (within ≈ 0.5 ps) as the intra- or inter-valley electron relaxations and the slow decay as carrier recombination and thermal cooling. In addition to the transient carrier dynamics, coherent optical or zone-edge acoustic phonons generated by the ultrashort pulses can contribute to the ΔT signal.[35, 36] Upon applying the high-frequency data filtering, as separately shown as the red line in Fig. 2a, we can clearly observe the high-frequency modulation of the ΔT signal as induced by phonon oscillation. Fourier-transform (FT) of the time-domain oscillation in Fig. 2b reveals two dominant CP peaks, one at ≈ 4.17 THz and the other at ≈ 7.26 THz.

In order to clarify the origin of these CP signals, we performed DFT calculations within the PBEsol exchange-correlation functional[37] and obtained the phonon dispersion of monolayer MoSe$_2$. While substrate and encapsulating layers can possibly affect the phonon character of 2D materials by renormalizing their electronic band structure and/or imposing additional restoring forces,[38, 39] modeling dielectric environments in a realistic fashion is beyond the scope of this work.[40, 41] Based on the calculated phonon dispersion shown in Fig. 2c and previous Raman measurement data,[42] we assign the signal at ≈ 7.26 THz to the optical A$_1$'(Γ) phonon, of which the out-of-plane atomic motion is schematically illustrated in the inset. Because the photon momentum is negligible in comparison to the crystal momentum, first-order Raman scattering will exclusively create Γ-point phonons. Within the phonon dispersion of Fig. 2c, however, such Γ-point phonon modes are missing for the ≈ 4.17 THz signal. Instead, the ZA phonons at the K- and M-points are close to the measured signal in terms of phonon energy. Based on the constraint of momentum conservation during the transfer of an electron between K and K' valleys in monolayer MoSe$_2$, we identify the ≈ 4.17 THz signal as the zone-corner phonon mode, ZA(K), of which the atomic motion is presented in Fig. 2d.

Although the momentum required for K-point phonon generation can be in principle provided by crystal defects, we can exclude this possibility because physical or chemical defects are scarce in the h-BN encapsulated TMDC monolayers. The high quality of our monolayer MoSe$_2$ sample was separately confirmed by negligible defect-related signals in the photoluminescence (see Fig. S1, ESI†). Another strong evidence that defects can be excluded as the source of the 4.17 THz signal is that the bare monolayer MoSe$_2$ without h-BN encapsulation, in which a large defect density was identified from photoluminescence

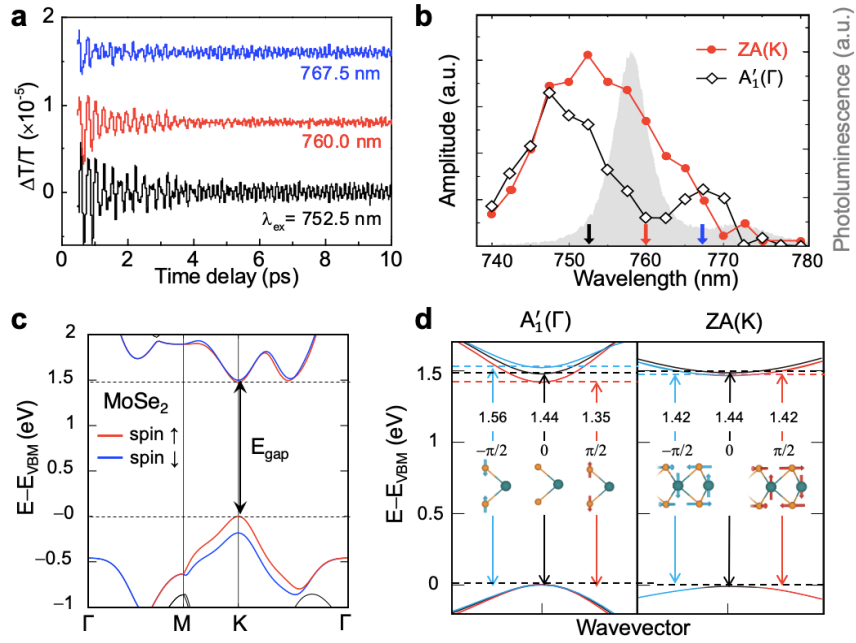

**Fig. 3** Excitation energy dependence of the $A_1'$ ($\Gamma$) and ZA(K) mode in MoSe$_2$ monolayer. (a) Coherent phonon oscillations at different excitation wavelengths. The wavelengths of 767.5 nm, 760 nm, and 752.5 nm are, respectively, below, on, and above the A-exciton resonance of monolayer MoSe$_2$. (b) Integrated phonon amplitudes of the $A_1'$ ($\Gamma$) and ZA(K) mode as a function of excitation energy. The shaded region represents the photoluminescence as a reference, and the colored arrows indicate the wavelengths in (a). (c) Band structure of monolayer MoSe$_2$ obtained by the PBEsol DFT calculation. d Energy gap modulation of monolayer MoSe$_2$ accompanied by the $A_1'$ ($\Gamma$) and ZA(K) phonon modes in three vibrational phases.

measurements, had a comparable strength of the 4.17 THz CP signal (see Fig. S2, ESI†).

To reinforce the activation of the ZA(K) mode in the CP measurement of MoSe$_2$ monolayer, we additionally obtained the $\Delta T$ data by simultaneously changing the pump and probe energy around the A-exciton. As mentioned earlier, the strong photoluminescence in monolayer TMDCs makes it difficult to study the phonon dynamics near the A-exciton energy with conventional Raman measurements. In contrast, CP measurements are free from this limitation, and the signal can be resonantly enhanced either in the generation or detection process. Fig. 3a shows CP oscillations ($\Delta T/T$) at three different excitation energies, and the corresponding FT spectra confirm the activation of both the $A_1'(\Gamma)$ mode at 7.26 THz and ZA(K) at 4.17 THz. Fig. 3b shows the excitation-energy dependence of the integrated phonon amplitudes for the two phonon modes, which were obtained by calculating the root mean square of the integrated FT intensity of each phonon mode and clearly exhibits resonance enhancement near the A-exciton. We previously observed a double peak-shaped FT amplitude of $A_1(\Gamma)$ phonon mode in the CP measurement of monolayer WSe$_2$, and attributed it to the phonon-induced energy-gap modulation.[31] To explain the microscopic origins in detail, we show in Fig. 3c the DFT-calculated electronic energy band of monolayer MoSe$_2$, and in the left panel of Fig. 3d we show its modulation according to the atomic motion of the $A_1(\Gamma)$ phonon. Examining the three vibrational phases, $\phi$ = -$\pi$/2, 0 and +$\pi$/2, calculated at one phonon per formula unit (see Fig. S3, ESI†), we find that the $A_1'(\Gamma)$ phonon induces a bandgap increase of ≈ 0.12 eV at $\phi$ = -$\pi$/2 and a decrease of ≈ 0.09 eV at $\phi$ = +$\pi$/2. These bandgap modulations should then reduce the integrated FT amplitude near the original A-exciton resonance energy. This energy-gap modulation by the $A_1(\Gamma)$ phonon is indeed confirmed by our optical measurement showing the $\pi$-phase shift of the $\Delta T/T$ oscillation at the two peak energies (see Fig. S4, ESI†).

In contrast, note that the 4.17 THz signal in Fig. 3b is characterized by a single peak in the wavelength dependence. Its assignment to the ZA(K) mode is then consistently justified by DFT calculations, which, as shown in the right panel of Fig. 3d, produce symmetric bandgap increases for both the $\phi$ = +$\pi$/2 and $\phi$ = -$\pi$/2 phases of the ZA(K) motion. Importantly, compared with the $A_1'(\Gamma)$ phonon, the bandgap change due to the ZA(K) mode is marginal (≈ 0.02 eV), and supports the single peak near the A-exciton resonance energy in the integrated FT amplitude. Upon closer inspection, one can also notice that the peak location (752 nm or 1.648 eV) of ZA(K) amplitude does not coincide with the A-exciton energy (758 nm or 1.636 eV), but is higher by approximately the ZA(K) phonon energy (0.017 eV).

Overall, we can consistently construct the valley dynamics picture as summarized below. First, remind that the inter-valley scattering mechanisms in semiconducting TMDC monolayers are dictated by the electron-spin configurations around the K and K' valleys or the strong spin-valley coupling. The conduction band edges of monolayer MoSe$_2$ are spin-split by ≈ 20 meV,[43] and the low-energy split K band will be occupied under the resonant excitation condition [Fig. 1(c)]. Next, an electron at the K valley excited with the photon energy of $E_A+E_{ZA(K)}$ can scatter into the K' valley by emitting one ZA(K) phonon. Note that the electron spin then should be flipped during the K-to-K' inter-valley scatterings since the spin configuration at K valleys is opposite to that at K' valleys. The spin-invariant inter-valley scatterings are less likely to occur because a final state at K' is energetically higher than an initial state at K. The requirement of spin-flip process explains why only the ZA(K) phonon, not LA(K) or TA(K) phonons, are excited in monolayer MoSe$_2$. According to the Eliott-Yafet spin relaxation mechanism,[44, 45] the spin-flip is possible through the carrier-phonon scatterings that will perturb the SOC. However, the Eliott-Yafet mechanism also dictates that a phonon mode in the spin-flip scattering must be odd-parity under the mirror-symmetry operation.[46] As shown in the inset of Fig. 2d, the ZA(K) mode indeed possesses odd-parity mirror symmetry with respect to the basal plane while other zone-corner acoustic phonons of LA(K) and TA(K) have even-parity under mirror operation and cannot induce the required spin flip (see Fig. S5, ESI†).

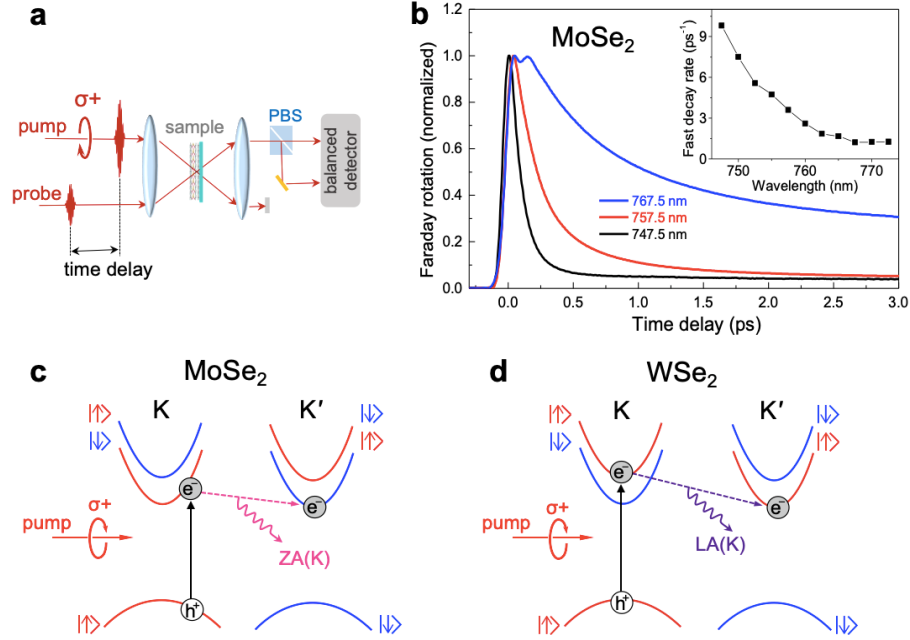

**Fig. 4** Faraday rotation in MoSe$_2$ monolayer and phonon mediated inter-valley scattering mechanisms. (a) Experimental setup for the Faraday rotation measurements. (b) Time-resolved Faraday rotations created by circularly polarized pump pulses at excitation wavelengths of 767.5, 757.5, and 747.5 nm. The inset shows the valley depolarization rate as a function of excitation wavelength. (c) Schematic of the inter-valley scattering of the photo-excited carriers with the generation of ZA(K) phonons in monolayer MoSe$_2$. Spin is flipped during the valley transfer. (d) Schematic of the inter-valley scattering mechanism with the generation of LA(K) phonons in monolayer WSe$_2$. Spin is preserved during the valley transfer.

The valley depolarization and the inter-valley electron scattering, which will concur with the ZA(K) phonon generation, can be identified by performing time-resolved Faraday rotation (FR) measurement schematically described in Fig. 4a. The FR signal provides the temporal evolution of the valley imbalance, which is created by a circularly polarized pump pulse. The FRs at different wavelengths in Fig. 4b indicate bi-exponential signal decay, and we focus on the fast or dominant depolarization process. When the excitation energy is below the A-exciton energy of 1.636 eV (758 nm), the valley depolarization is rather slow with a decay time of ≈ 0.80 ps. But, when the excitation energy is tuned to 1.653 eV (750 nm), which is approximately the sum of A-exciton energy $E_A$ and ZA(K) phonon energy $E_{ZA(K)}$, the valley depolarization becomes as fast as 0.18 ps. We note that the overall monotonic increase of the decay rate at shorter wavelength is consistent with previous reports, which demonstrated the accelerated valley depolarization with the excitation photon energy from the circular polarized photoluminescence degree measurements.[24, 27] This energy dependence indicates the acceleration of valley depolarization as the result of ZA(K) phonon-mediated inter-valley scatterings. Namely, the close correlation between the ZA(K) phonon amplitude and the valley-depolarization rate supports our conclusion summarized in Fig. 4c, or the carrier–phonon scattering involving the zone-corner phonon mode ZA(K) can become the primary channel for ultrafast valley depolarization processes in monolayer MoSe$_2$.

In order to reconfirm and further extend the valley depolarization mechanisms involving zone-corner phonons, we repeated the time-resolved CP and FR measurements with hBN-encapsulated monolayer WSe$_2$. Firstly, from the time-resolved FR measurements, we could find that significant valley depolarizations occur within a phonon period of 200 fs in monolayer WSe$_2$ (see Fig. S6, ESI†). As illustrated in Fig. 4d, the spin alignment of WSe$_2$ monolayer is opposite from that of MoSe$_2$ monolayer (see also Fig. S7, ESI†),[43] which allows the photoexcitation of higher branch conduction bands. With this energy-band configuration, the spin-maintaining inter-valley scattering between K and K' valleys, as denoted with the purple dashed arrow in Fig. 4d, is energetically more favorable. Accordingly, it is expected that the scattering in WSe$_2$ monolayer will involve K-point phonons other than the spin-flipping ZA(K) mode.

Fig. 5a displays the time-resolved ΔT signal at the A-exciton resonance for monolayer WSe$_2$. The FT spectrum of the CP oscillation in Fig. 5b reveals three phonon modes. Comparing again with the DFT-calculated phonon dispersion

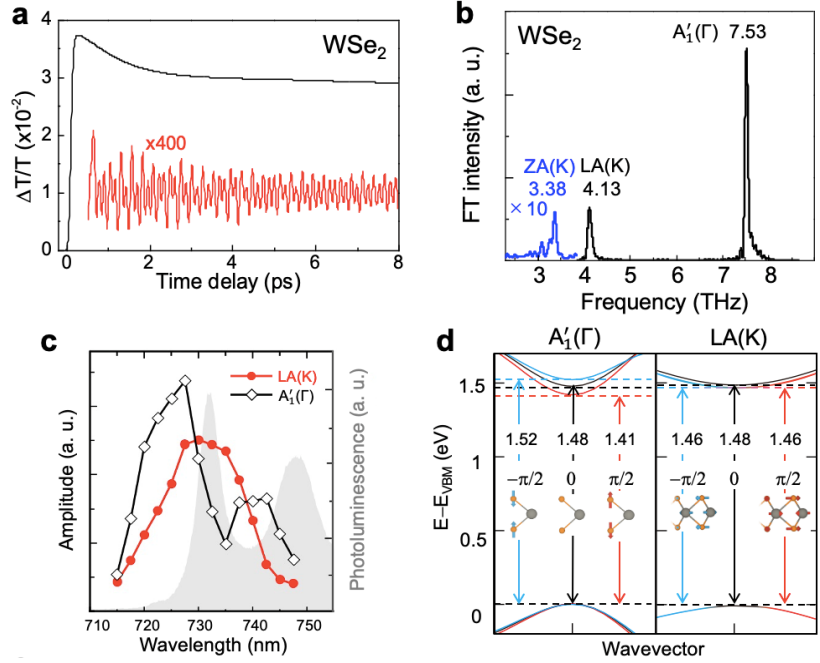

**Fig. 5** Coherent phonon oscillations in WSe$_2$ monolayer and excitation energy dependence. (a) Time-resolved transmission change and extracted ΔT oscillation in monolayer WSe$_2$ on the A-exciton resonance of 732 nm excitation. (b) FT spectrum of the ΔT oscillation revealing a strong LA(K) mode at 4.13 THz, as well as a A$_1$'(Γ) mode at 7.53 THz and a weak ZA(K) mode at 3.38 THz. (c) Phonon amplitudes of the A$_1$' (Γ) and LA(K) mode as a function of excitation wavelength. The shaded region represents the photoluminescence as a reference. (d) Calculated energy gap modulations at three vibrational phases as induced by the A$_1$' (Γ) and LA(K) modes.

(see Fig. S8, ESI†), we assign the signals at 7.53 THz, 4.13 THz, and 3.38 THz to the A$_1$'(Γ), LA(K), and ZA(K) modes, respectively. In our previous investigation of WSe2 monolayer,[31] we assigned a broad signal with its peak at 3.7 THz as the LA(M) mode. Such assignment was made based on the data extracted from unprotected samples that contain a large amount of defects and by the comparison with the phonon bands available in the literature. Accordingly, the present assignment of the signal at 4.13 THz to the LA(K) mode should be more robust thanks to the narrow phonon linewidth resulting from marginal defects in encapsulated WSe$_2$ monolayers and the comparison with our own phonon dispersion calculations (Fig. S8, ESI†). In this case, the LA(K) mode with even mirror symmetry is an order of magnitude stronger than the ZA(K) mode. We note that this observation in WSe$_2$ monolayer that contrasts the large ZA(K) generation in MoSe$_2$ monolayer is consistent with the proposed phonon-mediated valley depolarization model, where the participating phonon mode is determined by whether the electron spin is flipped during the inter-valley transfer. Signals from the TA(K) mode is not observed although it has even symmetry like LA(K), and we propose that it results from its electron-phonon scattering rate much weaker than that of LA(K).[47, 48]

Repeating the excitation energy dependent measurements on WSe$_2$ monolayer, as summarized in Fig. 5c, we find that the A$_1$'(Γ) mode exhibits a double-peak shape with energy,[31] but the LA(K) phonon amplitude features a single peak at the A-exciton resonance. In the same manner as the A$_1$'(Γ) and ZA(K) phonons in the monolayer MoSe$_2$ case (Fig. 3d), these observations can be understood in terms of the bandgap modulation shown in Fig. 5d that correspond to the A$_1$'(Γ) and LA(K) phonon modes. In spite of the overall similarity between the monolayer MoSe$_2$ ZA(K) and monolayer WSe$_2$ LA(K) signals in Fig. 3b and Fig. 5c, respectively, we also note a critical difference: The LA(K) in WSe$_2$ monolayer is peaked near the A-exciton, but the ZA(K) in MoSe$_2$ monolayer was resonant at the $E_A+E_{ZA(K)}$ energy condition. This discrepancy is again consistent with the different phonon-mediated inter-valley scattering mechanisms schematically shown in Figs. 4c and 4d for monolayer MoSe$_2$ and monolayer WSe$_2$, respectively.

## Conclusions

Based on mutually corroborating pump-probe experiments and DFT calculations, we demonstrated that valley depolarizations in semiconducting TMDC monolayers can be dominated by the K-point phonons. We showed that coherent zone-corner ZA(K) and LA(K) acoustic phonons are strongly activated in monolayer MoSe$_2$ and monolayer WSe$_2$, respectively, reflecting the strong spin-valley locking of electrons in the conduction band edge of two materials. The proposed valley depolarization channels produced by zone-corner phonon-mediated inter-valley scatterings were consistently supported by (i) the momentum compensation in the electron transfer from the K to K' valleys, (ii) the activation of only the ZA(K) mode with odd mirror symmetry in monolayer MoSe$_2$, and (iii) different shapes of the integrated phonon amplitudes for the ZA(K) [LA(K)] and A$_1$' (Γ) modes in monolayer MoSe$_2$ [WSe$_2$]. Our studies not only provide hitherto undisclosed inter-valley scattering mechanisms in semiconducting monolayer TMDCs but also suggest the intriguing possibility of controlling coherent phonon generations by exploiting the valley degree of freedom through a judicious two-dimensional material selection process.

## Methods

**Sample preparation.** A High-quality pristine sample is prepared by encapsulating a $MoSe_2$ monolayer with h-BN flakes at top and bottom sides. First, an h-BN flake with a thickness of ~100 nm is mechanically exfoliated and transferred onto a quartz substrate. Then, a monolayer $WSe_2$ exfoliated on a polymer film of poly methyl methacrylate (PMMA)-poly 4-styrensulfonic acid solution (PSS) is transferred to top of the pre-located h-BN flake using the dry transfer method. After this $MoSe_2$ monolayer transfer, we remove the PMMA film in acetone and annealing is performed in a mixture gas of Ar and $H_2$ to remove residual polymers. Finally, a top h-BN layer is transferred onto the monolayer $WSe_2$, yielding h-BN/$MoSe_2$/h-BN structure on quartz substrate.

**Phonon dispersion calculation.** Phonon dispersion is calculated with density functional theory as implemented in VASP package. The generalized gradient approximation (GGA) was used for the exchange-correlation energy within the PBEsol functional.[37 37] The kinetic cutoff energy is set as 500 eV, and the first Brillouin zone is sampled by an 18×18×1 Monkhorst Pack k-points grid for the unit cell. We adopt a slab model with a 20 Å interlayer spacing to circumvent spurious interlayer interaction between periodic images. The optimized lattice constant of monolayer $MoSe_2$ is 3.27 Å, which is in good agreement with previous studies.[49] With the optimized unit cell structure, we generate a minimal set of displaced 6×6×1 jsupercells using Phonopy package,[50, 51] and the dynamical matrix is obtained from harmonic force constants calculated with the finite displace method. The nonanalytical term correction vicinity of the center of the Brillouin zone center has been included as a post-processing correction of the dynamical matrix calculated with Born effective charges obtained by density functional perturbation theory (DFPT).[52] For the simulation of energy gap modulation due to phonons, the energy gap of displaced systems by phonons are calculated in which atoms in 3×3×1 supercell are moved in each phonon direction. The spin orbit coupling was included in the calculations of the band structure of $MoSe_2$ and $WSe_2$ and their energy gap calculations.

## Acknowledgments

This work was supported by the Basic Research Program (2017R1A2B3009872, 2020R1A2C1008368, 2020R1A6A1A03047771), Nano-Material Technology Development Program (2016M3A7B4024133), Global Frontier Program (2013M3A6B1078881), Basic Research Lab Program (Grant 2020R1A4A2002806) of the National Research Foundation of Korea (NRF) funded by the Korean government, and the Korea Institute for Advancement of Technology (KIAT) grant funded by the Korea Government (P0008458). S.B. was additionally supported by the Iwaki Scholarship Foundation of Japan.

## Conflicts of interest

There are no conflicts to declare

## Notes and references